%% file: Hlay_v3.tex
\documentclass[12pt]{article}
\usepackage{amsmath,amstext,amsfonts,amsbsy,amssymb,amscd,bbm,epsfig}
\usepackage{graphicx}
\usepackage{subfigure}
\input{macros.tex}

\usepackage{cprform}
=0cm \oddsidemargin=0cm

\newcommand{\be}{\begin{equation}}
\newcommand{\ee}{\end{equation}}

\newcommand{\f}{\frac}

\begin{document}
\begin{titlepage}
\begin{flushright}
ROM2F/2004/13
\end{flushright}
\vskip2truecm

\begin{center}
\begin{large}
{\bf  Layered Higgs Phase as a Possible Field Localisation on a Brane }

\end{large}
\vskip1truecm

P.~Dimopoulos$^{(a)}$\footnote{E-mail: dimopoulos@roma2.infn.it} and
K.~Farakos$^{(b)}$\footnote{E-mail: kfarakos@central.ntua.gr}



\vskip1truecm

{\sl $^{(a)}$ INFN-Rome2 Universita di Roma 'Tor Vergata' \\
Dipartimento di Fisica I-00133, Rome, Italy\\}

{\sl $^{(b)}$ Physics Department, National Technical University\\
15780 Zografou Campus, Athens, Greece}

\end{center}
\vskip1truecm

\begin{abstract}
\noindent So far it has been found by using lattice techniques that in the 
anisotropic five--dimensional 
Abelian Higgs model,  a layered Higgs phase exists in addition to the expected  
five--dimensional one. The exploration of the phase diagram has shown that the two 
Higgs phases are separated by a phase transition from the confining phase. This transition is  known to be 
first order. In this paper we explore the
possibility of finding a second order transition point in the critical line which separates the 
first order phase transition from the crossover region. This is shown to be the case only for the four--dimensional 
Higgs layered phase whilst the phase transition to the five--dimensional broken phase remains first order.
The layered phase serves as the possible realisation of four--dimensional spacetime dynamics which is embedded
in a five--dimensional spacetime. These results are due to gauge and scalar field localisation by 
confining interactions along the extra fifth direction.
\end{abstract}
\end{titlepage}

\section{Introduction - Motivation}

Since the mid eighties   lattice gauge models with
anisotropic couplings  defined in higher D-dimensional spaces  have been proposed. These models 
may exhibit, through a phase transition,  a phase which is coulombic in (D-1)
dimensions and shows confinement along the remaining  dimension. 
In fact, this was the result  of Fu and Nielsen 
using mean field techniques in a five-dimensional pure U(1) gauge theory with anisotropic couplings 
\cite{funiel}. This new phase was called layered. 

The Monte--Carlo analysis which followed \cite{rab} supported the mean field
results and helped to get a more precise picture of the phase diagram
\cite{stam}. Also in \cite{dim1} the orders of the phase transitions have been
analysed \footnote{It has to be noticed that for non--Abelian gauge theories the layer phase exists 
in six dimensions \cite{funiel,rab}. For the lattice realisation of the four--dimensional
confining phase in a five--dimensional non--abelian gauge theory in the context of 
a {\it compactified} extra dimension the reader may refer to \cite{ejiri, farakos}.}.

In addition, as  it may have been expected, the consideration of the interaction
with a scalar particle leads to a richer phase diagram. Actually, the
exploration of the phase diagram of the model for various sets of lattice
parameters values provides strong evidence that the layer phase is stable and it appears 
either in a   Higgs phase  for  the U(1) case  \cite{dim2, dim3}, or in a Coulomb phase for a  SU(2) adjoint Higgs model 
\footnote{Recently a paper appeared \cite{rummu} which presents a non--perturbative study of the Dvali--Shifman 
mechanism \cite{dvali} of the gauge localisation on a brane. For that reason a  SU(2) gauge theory with an adjoint 
scalar, whose mass parameter is space dependent, is employed in 3D.} 
\cite{dim4}. 

Since gauge theories defined on a $D >4$ spacetime are known to be non--renormalisable an explicit
cut--off $\Lambda$ has to be introduced \cite{dienes}. Therefore the theory is to be considered as an effective theory 
which emerges from a more fundamental renormalisable theory (for example the string theory). For the U(1)
gauge field the introduction of the cut--off $\Lambda$ leads to the admission of the strong coupling phase 
to be the  interesting phase for the five--dimensional theory. As a consequence the lattice methods have to be used 
as the unavoidable non--perturbative tool for the study of the system. 

Up to now the Monte--Carlo results  show that the transition
between the five dimensional strong coupling phase and the layered Higgs phase is
first order. A multilayer structure arises which 
supports the idea of the confinement along the extra dimension \cite{dim3, dim4}.
A crucial question may arise: is there any possibility for this phase transition to be 
of second order? We work on this possibility and we look for a second order ending point along the 
first order critical line \footnote{A similar behaviour has been seen in U(1)--Higgs model in 
4D \cite{alonso}, in SU(2)--Higgs model in 3D \cite{rummu_2} and in SU(2) adjoint Higgs model in 3D \cite{tepper}.}. 
This would  give evidence for the layer mechanism to be more realistic and useful in scenarios concerning the 
localisation of the fields on the four--dimensional subspace.

Before proceeding to the lattice model  let us 
present the action of the U(1)--Higgs model   
in five dimensions which in principle could inspire the lattice action used 
in the sequel for the numerical simulation. 

We assume  a five dimensional anti de Sitter space ($\mbox{AdS}_5$) with one  warped extra dimension. 
In general  the metric reads:
\begin{equation}
ds^2 = \alpha^2(z) [dx_{0}^2 - d\vec{x}^2] - dz^2
\end{equation}
We consider  $\eta_{\mu\nu}$ to be  the four dimensional Minkowski metric and $\alpha(z)$  the warp factor. 
We do not need to define  explicitly the form of the warp factor. We only require that it goes to zero 
as $z \rightarrow \infty$ (\cite{randall},\cite{kehagias}, \cite{giovannini}, \cite{Shaposh}). 
Hence the five--dimensional metric is written:
\begin{equation} \label{metric}
g^{MN} = (\displaystyle{\frac{1}{\alpha^2(z)}} \eta^{\mu \nu}, -1)
\end{equation}

We consider   now that  in such a space we define  a five dimensional Abelian Higgs model, the action of which reads: 
\begin{eqnarray}
S &=& S_{\mbox{gauge}} + S_{\mbox{scalar}} \nonumber \\
  &=& -\displaystyle{\frac{1}{4g_{5}^{2}} } \int d^5 x \sqrt{g}~~ F_{MN} F_{KL} g^{M K} g^{N L}
  + \int d^5 x \sqrt{g} ~~ \left [D_{M} \Phi^{*} D_{N} \Phi g^{MN}-V(\Phi)\right] \nonumber \\
  &=& \int d^4 x dz \left [-\displaystyle{\frac{1}{4g_{5}^{2}} } F_{\mu \nu} F_{\kappa \lambda} \eta^{\mu \kappa}
  \eta^{\nu \lambda} - \displaystyle{\frac{\alpha^2(z)}{2 g_{5}^{2}}} F_{\mu 5}F_{\nu 5} \eta^{\mu \nu} \right ] + 
  \nonumber \\
  &+& \int d^4 x dz \left [ \alpha^2(z) D_{\mu} \Phi^{*} D_{\nu} \Phi \eta^{\mu \nu} - \alpha^4(z) D_{z} \Phi^{*} 
  D_{z} \Phi - \alpha^4(z) V(\Phi) \right ] \label{action_c}
\end{eqnarray}
We note that  the upper case indices refer to the 5--D space, $M,N,K,L = 0,...,4$ and the lower case Greek ones to the 
4--D  space i.e. $\mu, \nu, \kappa, \lambda = 0,...,3$. It is obvious that the scalar field $\Phi$ depends on the 
five  dimensional space $(x,z)$.
Then we use the rescaling: 
$ \alpha(z) \Phi = \varphi $ for the scalar field. In the rather general case where the quartic scalar potential is 
considered, the scalar action takes the form:
\begin{equation}
S_{\mbox{scalar}} = \int d^4 x dz \left [ D_{\mu} \varphi^{*} D^{\mu} \varphi - \alpha^2(z) D_{z} \varphi^{*} 
D_{z} \varphi     
- M(z)^2 \varphi^{*} \varphi
- \lambda (\varphi^{*} \varphi)^{2} \right ] \label{scalar_c}
\end{equation}

\noindent where $M^2(z) = \alpha^2(z) m^2 + [\alpha^{\prime}(z)]^{ 2} + \frac{1}{2} [\alpha^2(z)]^{''} $ \footnote{
Assuming that $m^2 < 0$ on the brane ($z=0$), we note that 
depending from the exact form of the warp factor the mass term 
may turn to be positive after a certain distance  
or at least tends to zero asymptotically  along the transverse direction. 
So we meet the situation of  two degenarate minima near the brane and only one minimum far away from it.}.

It is a trivial matter for the action to be analytically continued to the Euclidean space from which the lattice action 
can be defined  after following  the usual methods for discretization. Therefore we take: 
\begin{eqnarray}
S_{\small L} &=& S_{\mbox{gauge}} + S_{\mbox{scalar}} \nonumber \\
&=& \beta_{g}  \sum_x\sum_{1 \le \mu<\nu \le 4}(1-\cos U_{\mu \nu}(x))
+  \sum_x\sum_{1 \le \mu \le 4} \beta_g^{\prime} ~ (1-\cos U_{\mu 5}(x))
\nonumber \\
&+& \beta_{h}  \sum _{x} \sum_{1 \le \mu \le 4} [ \varphi_{\small L}(x)
- U_{\hat \mu}(x) \varphi_{L} (x+a\hat \mu)]^{*}
[\varphi_{\small L}(x)
- U_{\hat \mu}(x) \varphi_{\small L} (x+a\hat \mu)]
\nonumber \\
&+& \beta_{h}^{\prime}  \sum _{x} [\varphi_L(x)
- U_{\hat 5}(x) \varphi_L (x+a\hat 5)]^{*}
[\varphi_{\small L}(x)
- U_{\hat 5}(x) \varphi_{\small L} (x+a\hat 5)]
\nonumber \\
&+&\sum _{x}m^{2}_{L}\varphi^{*}_{\small L}(x)\varphi_{\small L} (x)
+ \beta_{R}  (\varphi ^*_{\small L}(x)\varphi_{\small L} (x))^2], \label{action_l}
\end{eqnarray}


\noindent We denote  by $\varphi_{L} (x)$ the lattice scalar field and   
\begin{eqnarray}
U_{\mu\nu}(x) &=& U_\mu(x) U_{\nu}(x+a\hat\mu)U_{\mu}^{\dagger}(x+a\hat\nu)  U_\nu^{\dagger}(x) \nonumber \\ 
U_{\mu5}(x) &=& U_\mu(x) U_{5}(x+a\hat\mu)U_{\mu}^{\dagger}(x+a\hat5)  U_5^{\dagger}(x)  
\end{eqnarray}
are the plaquettes on the four--dimensional space  and along the fifth direction respectively 
The  U's are the links for the gauge field on the lattice \footnote{Notice also that here we use the symbol $x$ for 
the whole discretised five--dimensional space. The extra direction now is denoted by $x_T$.}. 
They are explicitly given by: $U_{M} = e^{i a A_{M}} ~~~(\mbox{with}~~M=1,...,5)$. 
The primed couplings refer to the interactions along the extra dimension. Moreover as it can be 
noticed from the corresponding continuous action, the couplings obey 
certain relationships, which depend on the warp factor \footnote{For the transition from the continous to the 
lattice action we have assumed the following rescaling 
for the scalar field: $2^{1/2} a^{3/2} \varphi = \beta_{h} \varphi_{L}$}. Hence we have: 
\begin{equation}
 \beta_{g}^{'}  = \alpha^2(x_T) \beta_{g}, ~~~~ \beta_{h}^{'} = \alpha^2(x_T) \beta_{h}, ~~~~ 
 \lambda = \frac{4\beta_R a}{\beta_{h}^{2} }
 \label{couplings}
\end{equation}
\begin{equation}
a^2 M^2(x_T) = \frac{2}{\beta_h}m^{2}_{L}
\end{equation}
Therefore  due to the assumed form for the warp factor the interactions for both the gauge and scalar fields 
are strongly coupled along the extra direction.\\
Since a brane is defined as any three dimensional submanifold to which ordinary matter is trapped \cite{Rubakov}
so that  it can not escape to the bulk, a possible realisation of the trapping mechanism 
is to assume the existence of confinement along the extra dimension. On the lattice this situation can be realised using  
a lattice model with anisotropic couplings. This is sufficient to lead   to the formation of the 
layered phase through a phase transition. 
In our context we consider this layered phase on the lattice as a possible paradigm on how a localisation of the
fields, obeying to non-perturbative interactions, 
may be carried out on the brane due to confining interactions in the bulk.  

In this paper we study a simplified realisation of the lattice action given by Eq. (\ref{action_l}) below. This is  
inspired by  Eq. (\ref{couplings}) i.e. to set the fifth (transverse) direction couplings to a strong 
coupling regime  while we neglect the explicit role of the warp factor in the lattice action. Therefore the lattice action 
(which  leads  to the five dimensional Higgs model
in the naive continuum limit (\cite{dim2}, \cite{dim3}) )  reads in  standard notation:
\begin{eqnarray}
S_{\small L} &=& S_{\mbox{gauge}} + S_{\mbox{scalar}} \nonumber \\
&=& \beta_{g}  \sum_x\sum_{1 \le \mu<\nu \le 4}(1-\cos U_{\mu \nu}(x))
+  \sum_x\sum_{1 \le \mu \le 4} \beta_g^{\prime} ~ (1-\cos U_{\mu 5}(x))
\nonumber \\
&+& \beta_{h}  \sum _{x} {\rm Re} [4 \varphi^{*}_{\small L}(x)\varphi_{\small L} (x)
- \sum_{1 \le \mu \le 4} \varphi^{*}_{\small L}(x)U_{\hat \mu}(x) \varphi_{\small L} (x+a\hat \mu)]
\nonumber \\
&+&   \sum _{x} \beta_{h}^{\prime} ~ {\rm Re} [(\varphi^{*}_{\small L}(x)
\varphi_{\small L} (x) 
- \varphi^{*}_{\small L}(x)U_{\hat 5}(x) \varphi_{\small L} (x+a\hat 5)]
\nonumber \\
&+&\sum _{x}[(1-2\beta_{R}-4 \beta_{h}- \beta_{h}^{\prime})\varphi^{*}_{\small L}(x)\varphi_{\small L} (x)
+ \beta_{R}  (\varphi ^*_{\small L}(x)\varphi_{\small L} (x))^2], \label{action_l}
\end{eqnarray}
Apart from the resulting simplicity in the context  of the phase diagram analysis,  
a connection of this work with previous studies of the layered phase can be achieved. Moreover our impression 
is that the full lattice model is likely to produce  physically similar results with the present simplified version.
This was also the case for the pure U(1) gauge model. 
The 'static' representation of the model for which  the gauge couplings were fixed by hand gave  
equivalent resuts with the model in which the warp factor was used for the scaling of the gauge couplings \cite{dim1}.

\section{The order parameters and the choice of couplings}
We study the abelian Higgs model on the lattice by using numerical methods.
The action is given explicitly by Eq. (\ref{action_l}). 
We define five order parameters, making also the distinction between space--like and transverse--like ones.
These are the following:

$$
\mbox{ Space--like Plaquette:} \hspace{0.3cm} P_S \equiv <\f{1}{6 N^5} \sum_x
\sum_{1 \le \mu<\nu \le 4} \cos U_{\mu \nu}(x)>
$$

$$\mbox{ Transverse-like~Plaquette:}
\hspace{0.3cm} P_T \equiv <\f{1}{4 N^5}
\sum_x \sum_{1 \le \mu \le 4} \cos U_{\mu 5}(x)>
$$

$$\mbox{Space-like~Link:}
\hspace{0.3cm} L_S \equiv <\f{1}{4 N^5} \sum_x
\sum_{1 \le \mu \le 4} \cos(\chi(x+\hat \mu) +A_{\hat \mu}(x)-\chi(x)) >
$$

$$\mbox{ Transverse-like~Link:}
\hspace{0.3cm} L_T \equiv <\f{1}{N^5} \sum_x
\cos(\chi(x+\hat 5) +A_{\hat 5}(x)-\chi(x))>
$$

$$\mbox{Higgs~field~measure~squared:}
\hspace{0.3cm} R^2 \equiv \f{1}{N^5} \sum_x \rho^2(x)
$$

\noindent We have assumed the polar form for the scalar field, i.e. $\varphi_{L} = \rho(x) e ^{i\chi(x)}$. 

\begin{figure}[!h]
\begin{center}
{\includegraphics[scale=0.4, angle=-0]{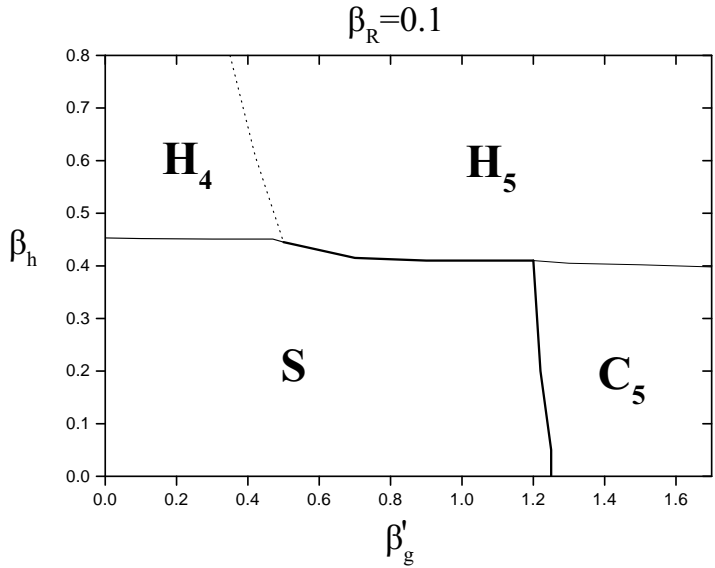}}
\caption{Phase diagram of the 5-D Abelian Higgs Model with the space--like gauge coupling set to 
the strong coupling $\beta_g=0.5$ (taken from \cite{dim3}). }
\label{phase_diag}
\end{center}
\end{figure}

In \cite{dim3} this model has been already studied and  a first exploration for
the phase diagram is available. In that work, since the parameter space is very large,
consisting of five lattice
parameters,the choice has been made to fix $\beta_{g}$ to 0.5, $\beta_{h}^{\prime}$ to 0.001 and
consider two values of $\beta_{R}$ (0.1 and 0.01) and explore the parameter space ($\beta_{g}^{\prime }$,
$\beta_{h}$). 
Under these conditions the analysis of the order parameters defined above yielded  a
phase diagram consisting of the three expected phases which are the
confining phase ($S$), the Coulomb phase ($C_5$) and the Higgs phase ($H_5$) each of them defined
in  five dimensions.  In addition  a fourth phase
is present: a Higgs phase in four dimensions ($H_4$) (see Fig.\ref{phase_diag}).
The distinction between
$H_4$ and $H_5$ can be achieved  due to the different behaviour of the transverse--like order parameters
within the two phases. Details and conclusions  on the existence of this layer Higgs phase
can be found in \cite{dim3}. Let us refer also that the identification for the order of the phase transitions
was possible and has lead to the conclusion that
(for the two values of $\beta_{R}$ used) both $H_4$,
$H_5$ are separated from the confining phase by a first order phase transition.  We reproduce the phase diagram 
for $\beta_{R}=0.1$ as it was depicted in \cite{dim3} (Fig.\ref{phase_diag}). 

\section{Searching for a second order phase transition}
At this point the question arises  whether it could be possible for the $H_4$ layer phase to appear via
a second order phase transition. 
Following \cite{dim3}, we consider the system being in the confining regime by setting
$\beta_g=0.5$ and fixing $\beta^{\prime}_{h}$ to the very small value 0.001. We expect 
the phase transitions to the  Higgs phases to be weaker as the Higgs self coupling $\beta_{R}$ increases. 
We explore the order of the $S-H_4$ phase transition 
by setting the transverse gauge coupling $\beta^{\prime}_{g}$ to 0.2 while we 
increase $\beta_{R}$. In advance, it should be noted that, 
as we move to larger values of $\beta_{R}$, the relative 
positions of the phases in the phase diagram  
are substantially similar to what is shown in  Fig.\ref{phase_diag} for $\beta_{R}=0.1$. 
So, setting 
$\beta^{\prime}_{g}$ to 0.2, we always explore the $S-H_4$ phase transition.

In the sequel we give strong evidence that, the
$S-H_4$ first order phase transition line ends at a second order point followed by a
crossover region. At the same moment the $S-H_5$ phase transition remains  first order. This 
additional fact confirms the special nature of the four dimensional layer Higgs phase. 

We give now  information for the simulating process. We used a 4--hit metropolis 
algorithm for the updating of the fields. In addition we implemented the global radial algorithm and the 
overrelaxation algorithm for the updating of the Higgs field. We used four lattice volumes, $ 8^5$, 
$10^5$, $12^5$, $14^5$, and we performed 20000--30000 measurements for each point which  we analyzed in the parameter space. 
We studied a large number of $\beta_{R}$ values
before concentrating our study to the interval $[0.140, 0.165]$ in which the first order phase transition 
turns to be a weaker one before it passes to the crossover region.

In the subsequent paragraphs we  present our results which are based upon  using the hysteresis loop
technique, the finite volume size scaling, the susceptibility  and
the study of the correlation functions for the Higgs field measure squared.

\subsection{Hysteresis loop technique results}

\begin{figure}[!h]
\begin{center}
{\includegraphics[scale=0.4, angle=-90]{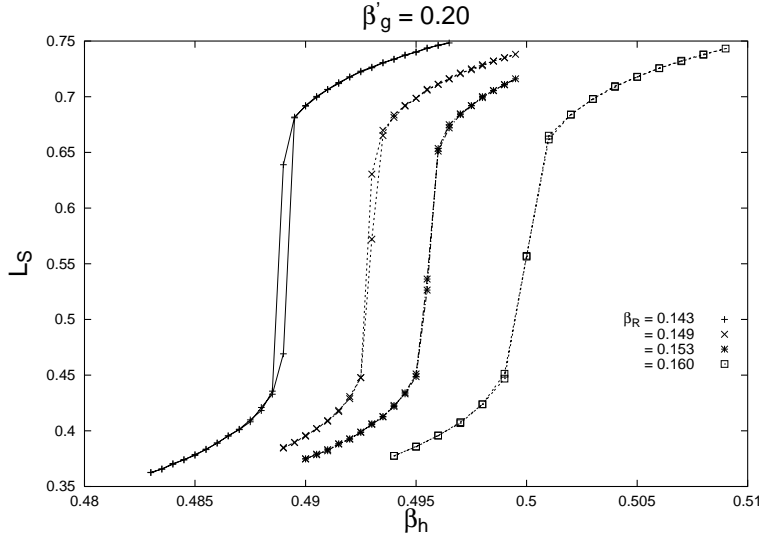}}
\caption{Hysteresis loops showing that the loop for the link--space order parameter 
disappears for $\beta_{R}$ values bigger than 0.153. }
\label{hyst_bgt020}
\end{center}
\end{figure}

The first tool for the exploration of the phase diagram with  $\beta_{R}$ 
is the hysteresis loop technique. 
Although this technique gives results that have to 
be taken into account with caution quantitatively, nevertheless they prove to be 
very useful as qualitative ones. To this end we use the hysteresis loop results as a general guide
to get a crude estimate on the $\beta_{R}$ interval within which the phase transition 
is converted from a first order  to a  higher order one. In Fig.\ref{hyst_bgt020} we depict the hysteresis 
loop results for the four --dimensional gauge invariant quantity $L_{S}$ 
and for four values of $\beta_{R}$, namely $\beta_{R}=0.143, 0.149, 0.153, 0.160$. The lattice 
volume in this example is $8^5$. One can see from the figure that while  
there is a well formed loop for $\beta_{R}=0.143$ indicating a first 
order phase transition, this changes to a smaller one 
for $\beta_{R}=0.149$, and it seems to  disappear for $\beta_{R}=0.153$. Although this value should not
be taken too seriously, one  should keep in mind 
that around the value $\beta_{R}=0.153$ a weaker phase transition is still present.
Furthermore we have to mention that the transverse link
quantity, $L_{T}$, (not shown in the figure) remains almost unaffected 
by the phase transition, being stuck to a very small 
value close to zero (for details see \cite{dim3}).
  
\begin{figure}[!h]
\begin{center}
\subfigure[]{\includegraphics[scale=0.25,angle=-90]{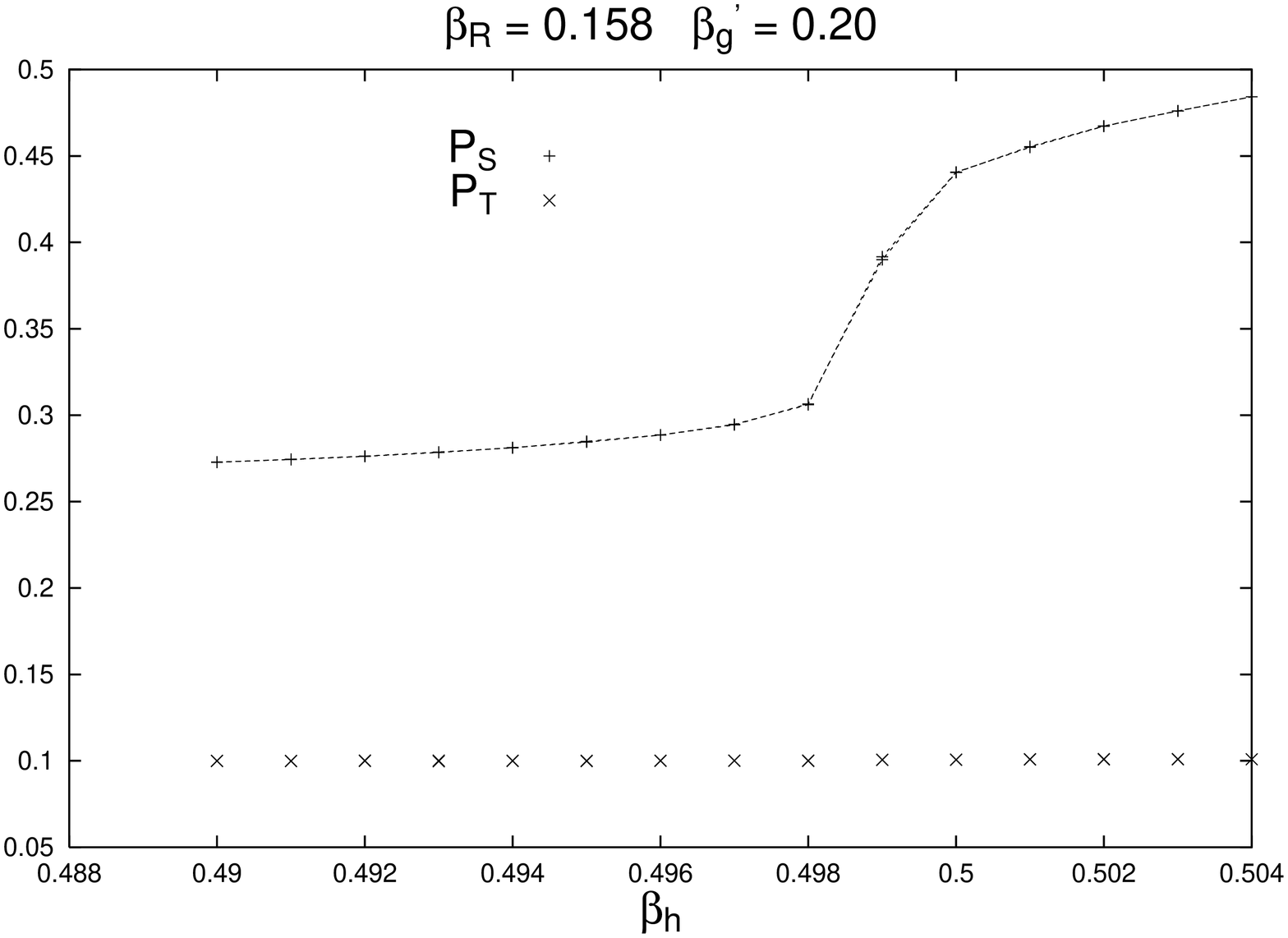}}
\subfigure[]{\includegraphics[scale=0.25,angle=-90]{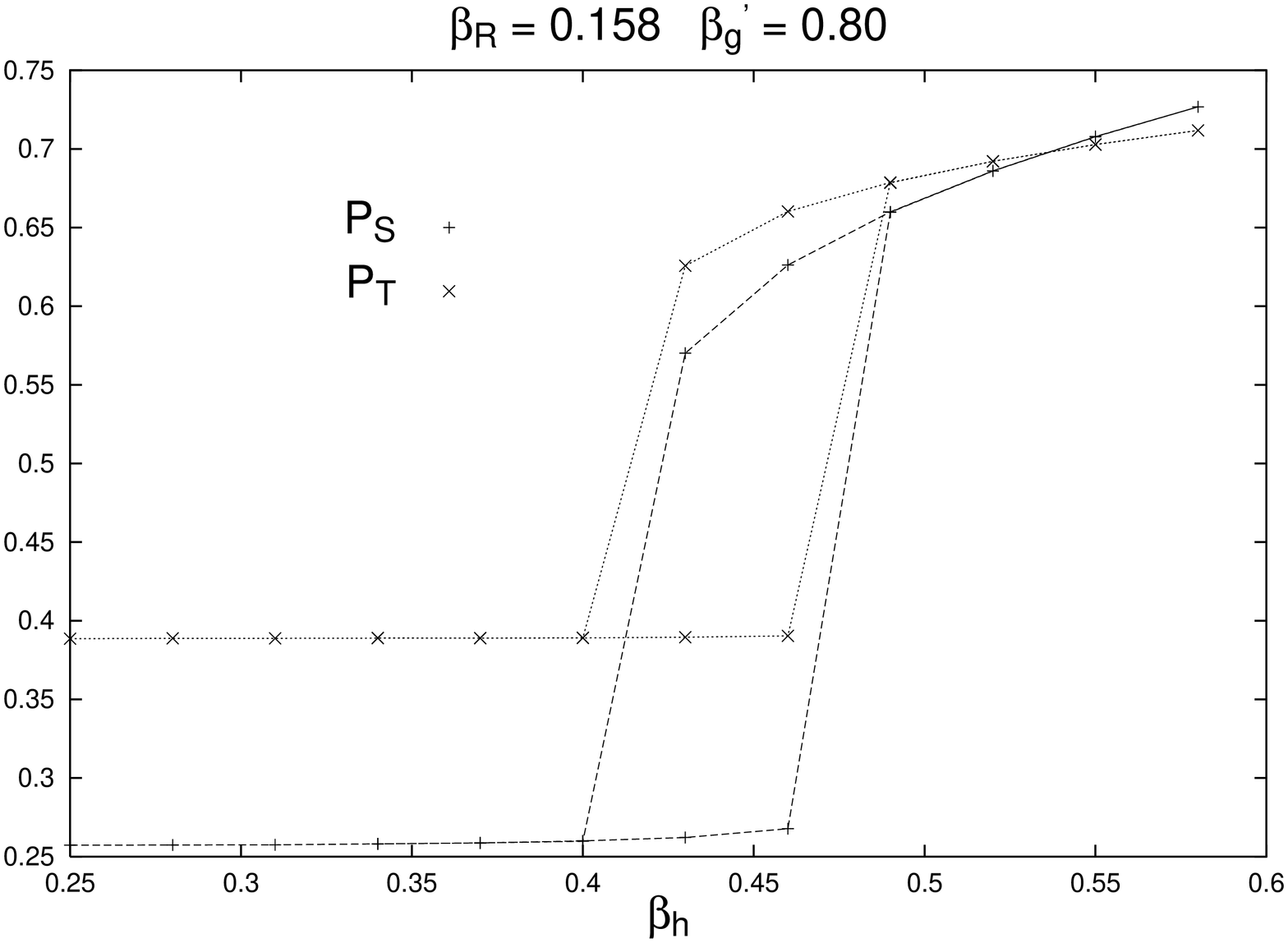}}
\subfigure[]{\includegraphics[scale=0.25,angle=-90]{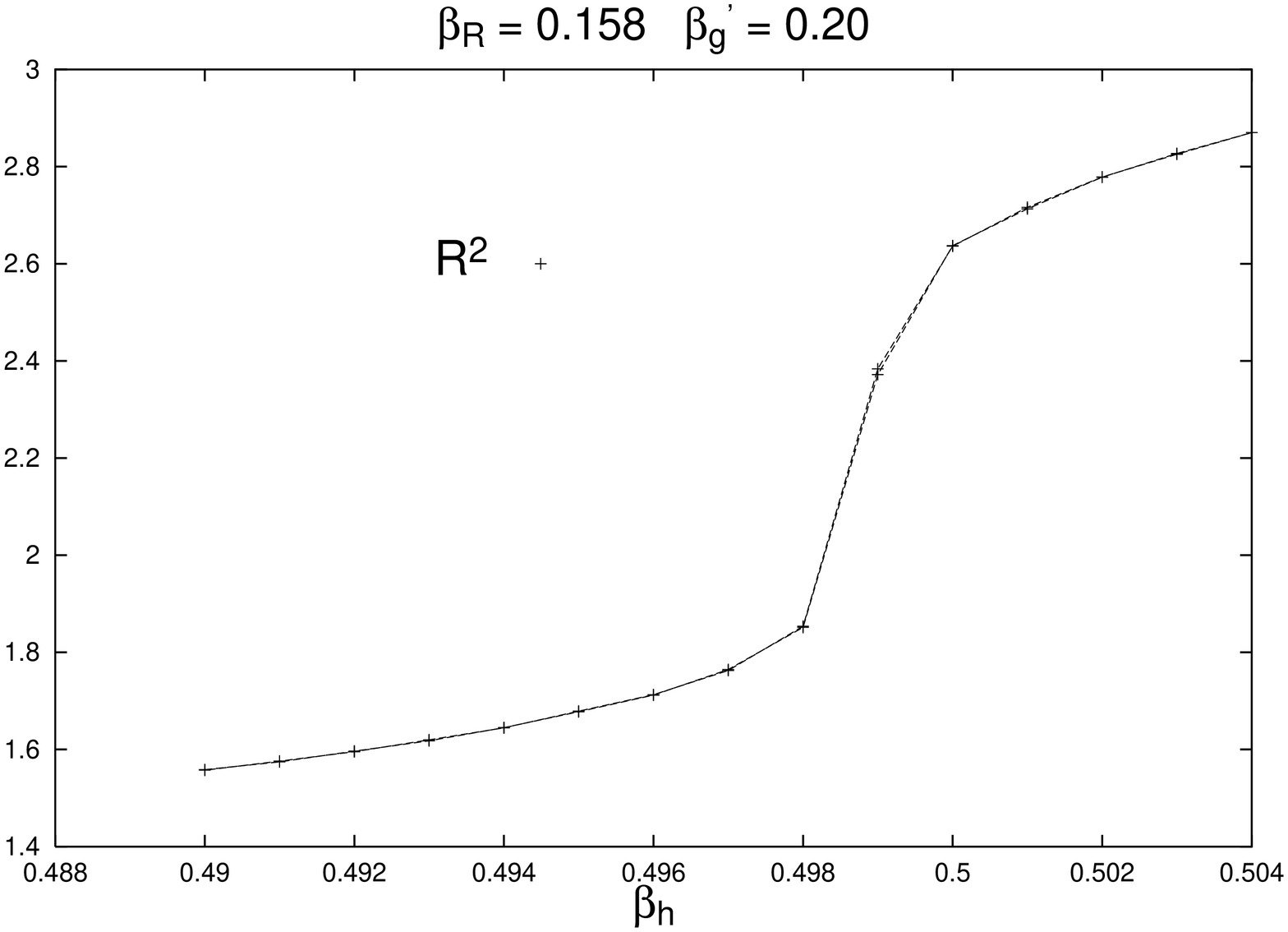}}
\subfigure[]{\includegraphics[scale=0.25,angle=-90]{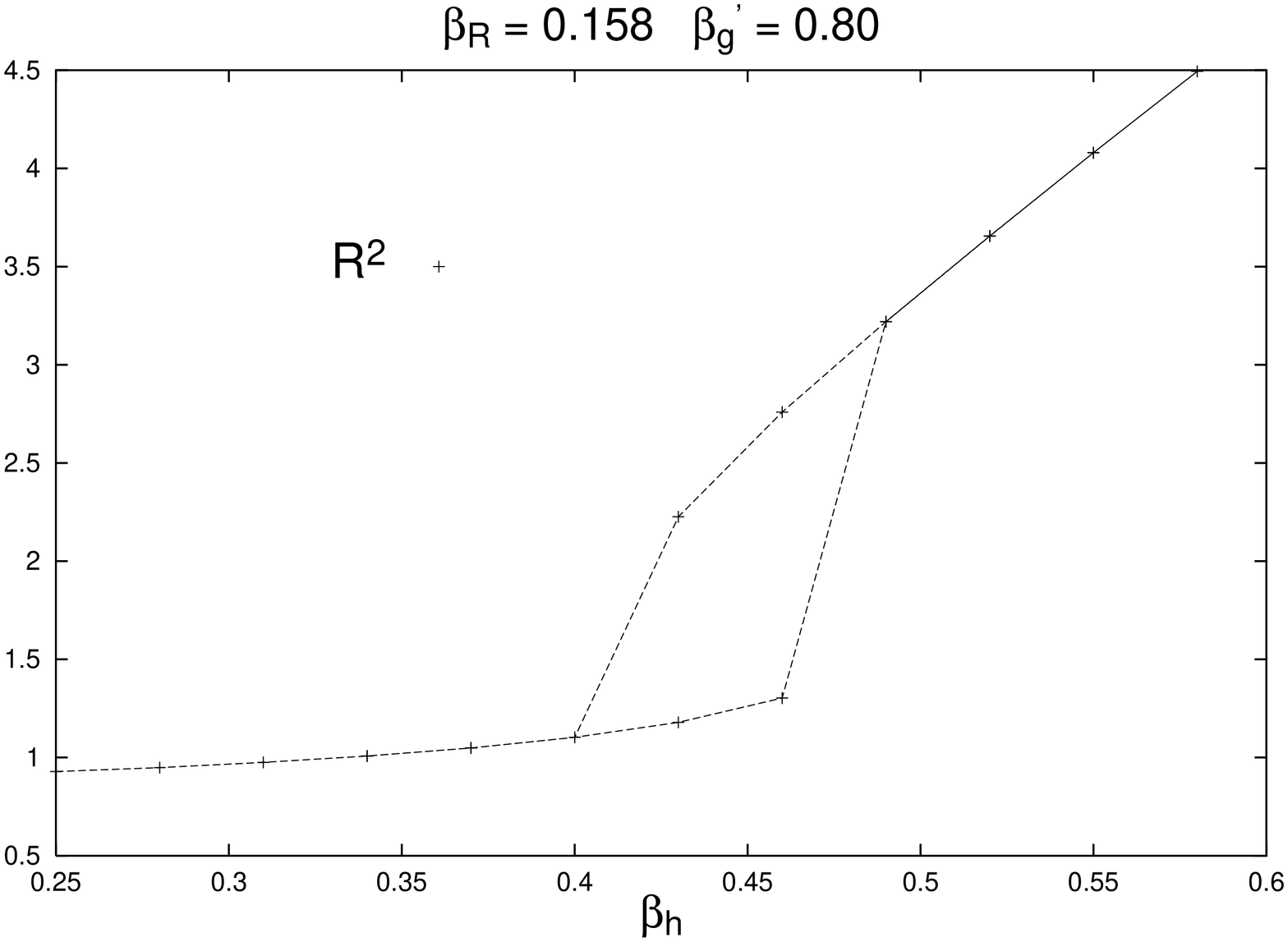}}
\caption{Characteristic examples showing the obviously different order of the phase transition 
for the cases $S-H_{4}$ and $S-H_5$ at  the same value of $\beta_{R}$, for two different values of 
$\beta_{g}^{\prime}$.}
\label{S_H4_H5}
\end{center}
\end{figure}

In  Fig. \ref{S_H4_H5} we give an example of the different phase transition orders of the
$S-H_4$ and $S-H_5$ transitions, both for  $\beta_{R}=0.158$ and lattice volume $8^5$. 
In Fig.\ref{S_H4_H5}a we present the hysteresis loop results 
on $P_{S}$ and $P_{T}$ for $\beta^{\prime}_{g}=0.20$. The behaviour of $P_{S}$ indicates a  
phase transition though a smooth one since there is no hysteresis loop, while the $P_{T}$
is almost constant and equals 0.1, in accord with the strong coupling prediction $\beta_{g}^{\prime}/2$. 
This figure should be compared with the Fig.\ref{S_H4_H5}b, which 
refers to  $\beta^{\prime}_{g}=0.80$. The hysteresis 
loop results shows a very strong first order phase transition, exhibited by both $P_{S}$, 
$P_{T}$\footnote{Notice that the unbroken phase is a confining one due to the 
fact that $P_{S}$ and $P_{T}$ follow the strong coupling limits $\beta_{g}/2$ and 
$\beta^{\prime}_{g}/2$ respectively (for more on that see \cite{dim3}).}. 
This behaviour refers to the $S-H_5$ phase transition. 
Figures  \ref{S_H4_H5}c and   \ref{S_H4_H5}d show the behaviour for $R^2$ which illustrates the fact 
that for both cases the system passes to a broken phase.
In other words, increasing $\beta{_g}^{\prime}$ one finds two different Higgs phases (see for example 
Fig.\ref{phase_diag}), a four--dimensional
and a five--dimensional one both  separated from the five--dimensional confining phase by 
phase transitions of different orders.

\subsection{Finite volume size scaling}
As it has been discussed  in \cite{dim3} one of the main features of the $S-H_4$ phase transition is the 
multi--layer structure. This means that since the system undergoes a transition 
to a four--dimensional phase rather than a five dimensional one, some special signal should appear.
Besides a first order phase transition this consists of a 
multipeak structure in the finite lattice volume histograms for the gauge invariant observables, 
instead of the  expected behaviour  of the two-peak structure. 
Furthermore it has been shown that every space--like gauge 
invariant quantity defined on each space--like volume (i.e. a four--dimensional layer) 'feels'
the phase transition for different pseudocritical values of the lattice parameters. Since this is
a consequence of the finite lattice volume used for Monte--Carlo simulations in combination 
with the four--dimensional dynamics when the layer phase arises, we justify the choice of 
analysing the results on the four--dimensional subspace. 

\begin{figure}[!h]
\begin{center}
{\includegraphics[scale=0.3, angle=-90]{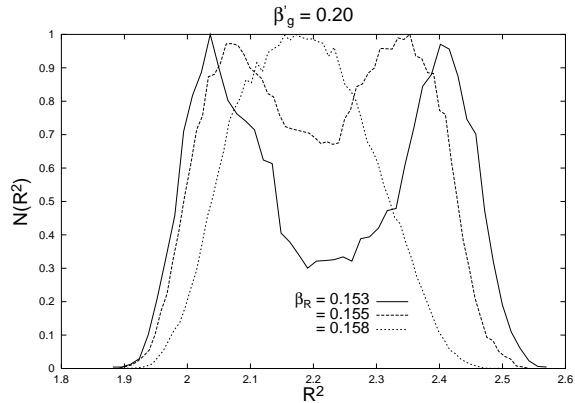}}
\caption{Histograms of $R^2$ over a space--like volume for three values of $\beta_R=0.153, 0.155, 0.158$
and lattice length $N=14$.}
\label{histo_br}
\end{center}
\end{figure}

\begin{figure}[!h]
\begin{center}
\subfigure[]{\includegraphics[scale=0.25,angle=-90]{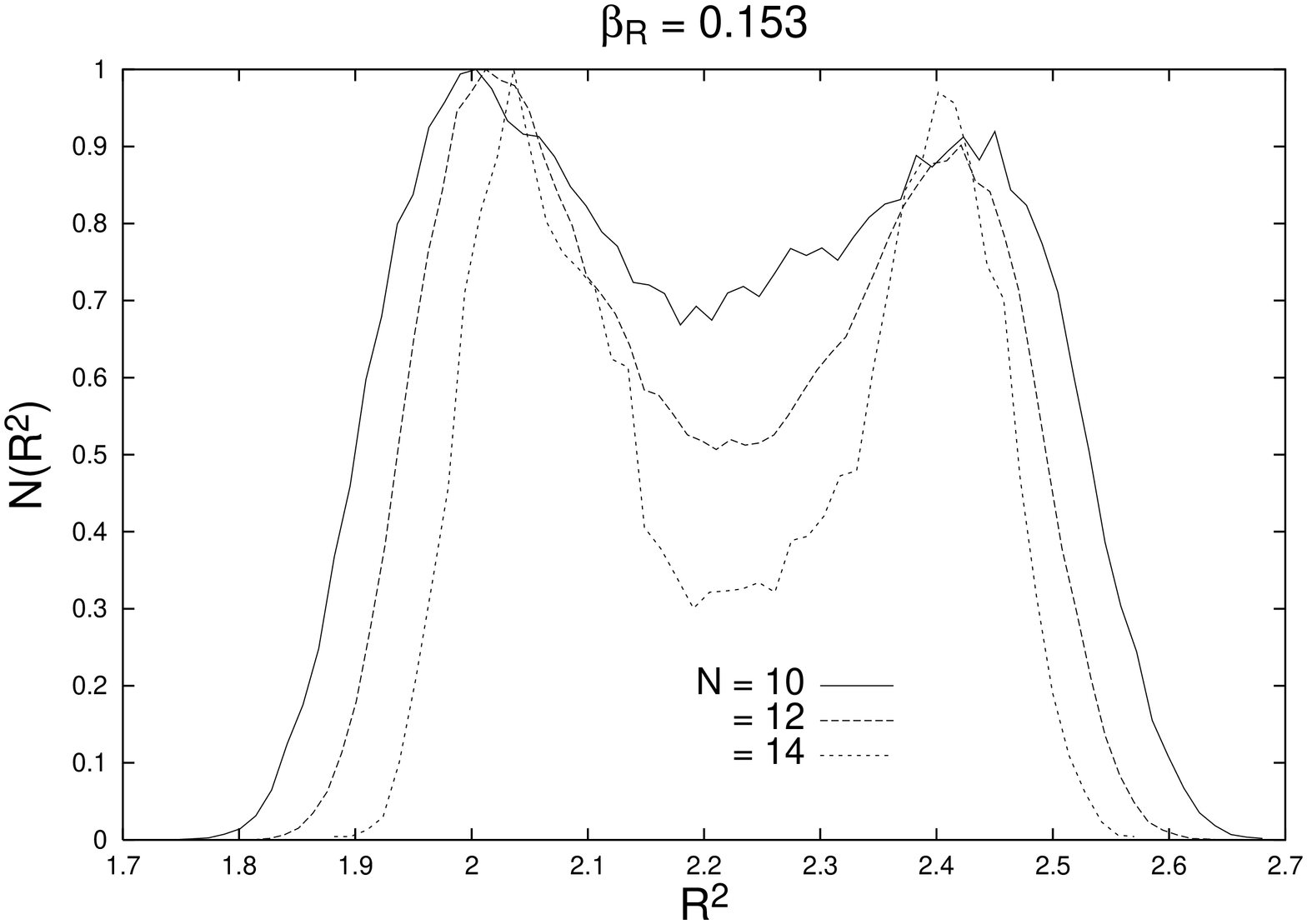}}
\subfigure[]{\includegraphics[scale=0.25, angle=-90]{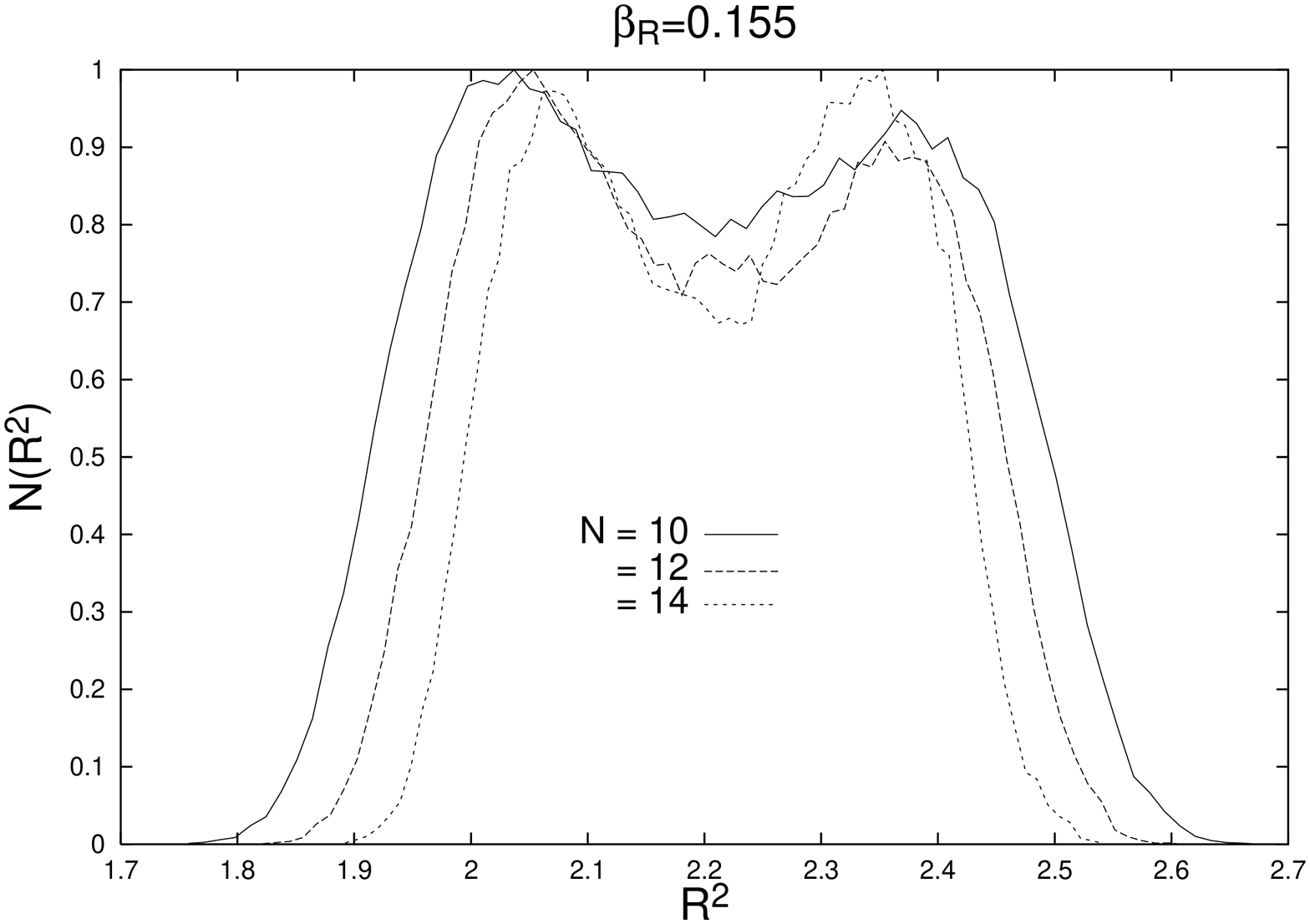}}
\subfigure[]{\includegraphics[scale=0.25,angle=-90]{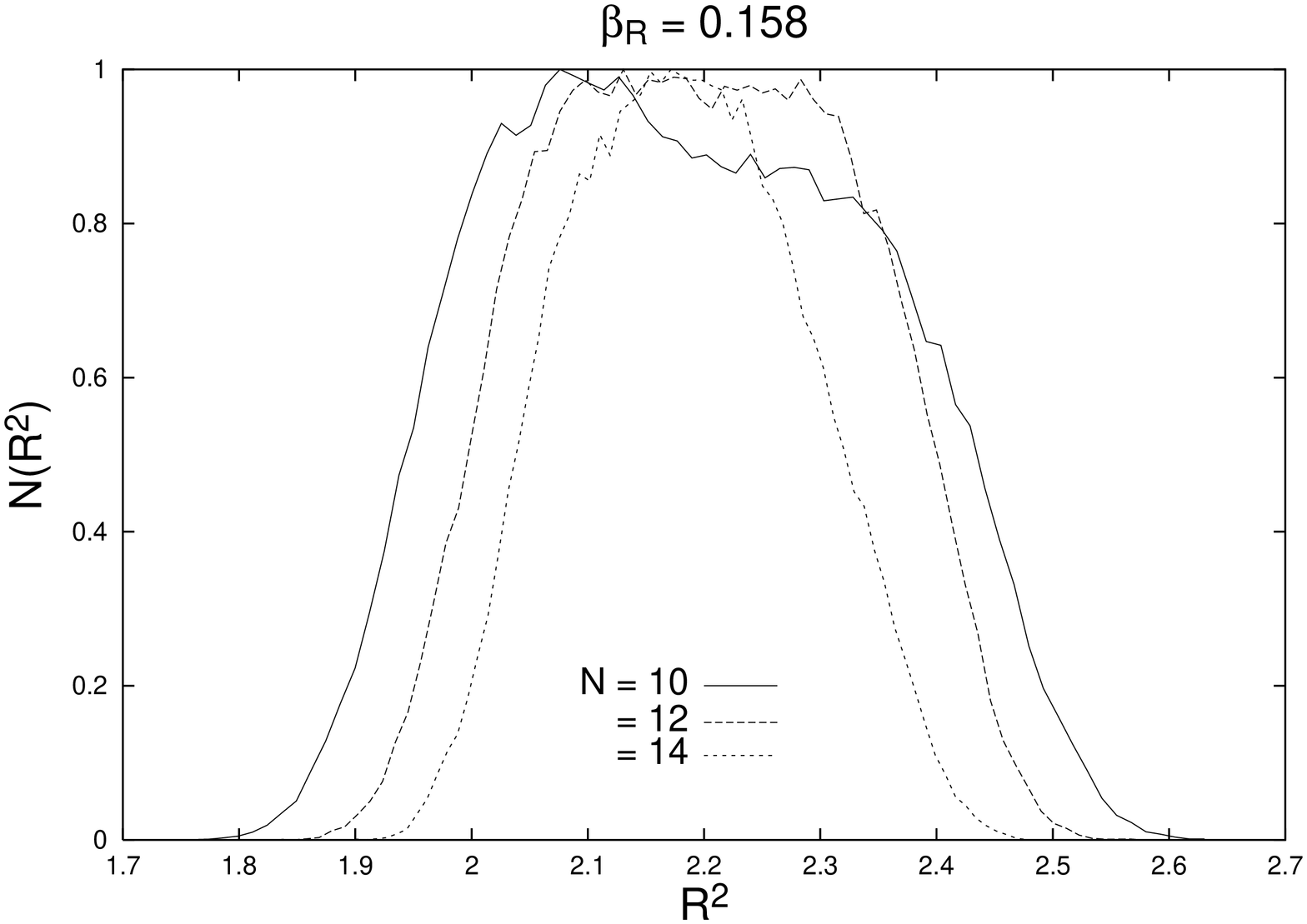}}
\caption{The histograms for $R^2$ as the lattice length increases for the three 
values of $\beta_R=0.153, 0.155, 0.158$.}
\label{histo_br_V10_12_14}
\end{center}
\end{figure}

In Fig.\ref{histo_br} we depict the histograms of the Higgs field measure squared, $R^2$ 
for $\beta^{\prime}_{g}=0.20$ and three values of 
$\beta_{R}$.
All the three histograms refer to   $\beta_h$ values in the critical region.
The lattice volume in this figure is $14^5$. The $R^2$ histograms 
refer to  four--dimensional (space--like) volume. The two peak structure is more
pronounced for the smaller value of $\beta_{R}$ (i.e. 0.153), where the two peaks are totally 
separated.
For $\beta_R=0.155$ the two peak structure is less emphasised while for $\beta_R=0.158$ it has 
already disappeared. 
In order for someone to use this method with more safety the lattice volume dependence of the two peak structure 
should be taken into account. 
This is provided  in Fig.\ref{histo_br_V10_12_14} . In Fig.\ref{histo_br_V10_12_14}a it is 
easily seen that the two peaks become well separated as the lattice length increases from $10$ to $14$ which
serves as an indication of a first order phase transition for the case of $\beta_{R}=0.153$. This has to be 
compared with the really inversed  behaviour for $\beta_{R}=0.158$ shown in Fig.\ref{histo_br_V10_12_14}c. 
The $\beta_{R}=0.155$
case, Fig.\ref{histo_br_V10_12_14}b, for which  the peak separation does not change significantly 
as the lattice length goes from $10$ to 
$14$, gives an estimate of a first order phase transition becoming much weaker and probably of higher order.

\begin{figure}[!h]
\begin{center}
{\includegraphics[scale=0.4, angle=-90]{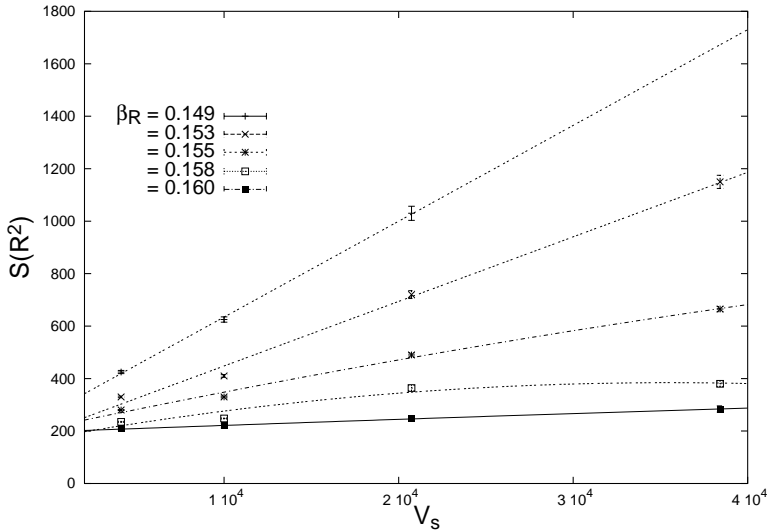}}
\caption{The susceptibility versus the space--like volume for five values of $\beta_R$ in the critical $\beta_R$ region.}
\label{susc}
\end{center}
\end{figure}

Let us now present more quantitative results by giving the results for the susceptibility 
of $R^2$ on the layers for various values of $\beta_{R}$. This is defined by:

$$ S(R^2)=V_s (<(R^2)^2>-<R^2>^{2}) $$

\noindent where $V_s$ denotes the space--like lattice volume. The results are depicted in Fig.\ref{susc}. 
The errors have been calculated by using the Jackknife method. It is known that a first order phase transition
is signalled by a linear increase of the maximum of the susceptibility with the volume. This is actually 
the case for $\beta_{R}=0.149 ~ \mbox{and} ~0.153$. The situation changes  for $\beta_{R}=0.155$ where 
the linear behaviour is apparently absent. In addition, for the bigger values $\beta_{R}=0.158 ~ \mbox{and} ~0.160$ there
is not a clear increase with the volume. This case corresponds to a crossover behaviour. 
Therefore, the conclusion is that in the vicinity of $\beta_{R} = 0.155$ we meet with the well known 
situation, where 
a first order phase transition line  ends to a second order phase transition point followed by a  
crossover.  

\subsection{Correlation functions}
In this section we present the behaviour of two  correlation functions, one  defined  on the 
whole five--dimensional space and the other  on the space--like, four--dimensional one. These correlation functions 
involve  the Higgs field measure squared $R^2$, defined in section 2. 
The definition of the correlation functions is given by:
\be \label{cor}
C_{S,T}(n)=\sum_{i}\frac{<(R^2)_i (R^2)_{i+n}> - <(R^2)_{i}>^2}{<(R^2)_i^2>-<(R^2)_{i}>^2}
\ee
where $n$ takes values from 1 to $N$ (i.e. the lattice size). The indices  $S$ and $T$ are used to 
distinguish the correlators. The one defined in  the transverse direction is noted with the index $T$. The other defined 
in the space--like volume is denoted with $S$.

\begin{figure}[!h]
\begin{center}
{\includegraphics[scale=0.4, angle=-90]{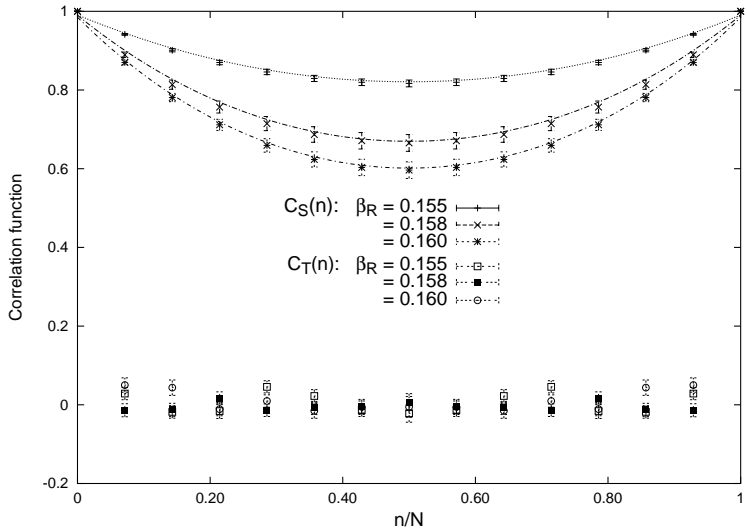}}
\caption{The space--like and time--like correlation functions for $L=14$ and for three values 
of $\beta_R = 0.155,~ 0.158,~ 0.160$, in the region of $\beta_h$ where the susceptibilities show a  peak.}
\label{cor_V14}
\end{center}
\end{figure}

The results for the two correlators are radically different. An example of our results is shown in 
Fig.\ref{cor_V14}. This refers to the case of $N=14$  lattice size for three values of $\beta_R$. 
We see that while $C_{T}$ decreases very fast, 
reaching zero and fluctuating around it, $C_{S}$ takes values  different from zero. 
This serves as a clear  evidence that a layered phase is formed. The layers are decoupled as a 
consequence of the strong coupling imposed on the transverse direction, which has the implication of 
vanishing $C_T$.
Moreover the rather reasonable behaviour of $C_{S}$ shows  that inside the layers a four dimensional 
dynamics is still met  as it might be expected.

\begin{table}[!h]
$$
\begin{array}{ccccccc}
\hline \\
\mbox{Lattice size} && \beta_R = 0.155 && \beta_R = 0.158 && \beta_R = 0.160
\\
\hline \\
10  && 0.145(3)  && 0.165(4) && 0.181(4)\\
12  && 0.112(3)  && 0.137(3) && 0.167(7)\\
14  && 0.090(3)  && 0.131(6)&& 0.152(7)\\
\hline \\
\end{array}
$$
\caption{The masses in lattice units. We observe that for $L=14$ and $\beta_R=0.155$ the value for the mass parameter has 
decreased by a factor of 1.7 in comparison with the $\beta_R=0.160$ corresponding value. }
\label{masses}
\end{table}

Another very interesting feature of the $C_S$ correlation function is that as the $\beta_R$ value 
decreases the 
curve becomes more flat. We should note that in the case of a second order phase transition and for infinite 
volume this should be really flat. 
This is a fact corresponding to infinite correlation length or vanishing mass 
for the lightest scalar mode. In other words, by adjusting the $\beta_h$ value into the critical region 
we might expect a mass behaviour of the type 
$m_s\propto (\beta_R - \beta_R^c)^{\nu}$.  The light scalar mass calculation can be achieved by using a fit of the 
form $const \times  \cosh(m_s(x-N/2))$ to the correlation functions $C_S$. The parameter $m_s$ is the dimensionless mass 
parameter of the scalar mode. An example of the fits is shown in Fig.\ref{cor_V14}. The results for $m_s$ for the cases 
considered are shown in Table \ref{masses}. From that Table  and for the largest lattice size used 
we can see that $m_s$ decreases by a factor of 1.7 between 
$\beta_R = 0.160$ and 0.155. A more clear signal for  the vanishing $m_s$ would require bigger volumes and still 
higher  computer time. Nevertheless, after considering the previous analysis on susceptibility 
combined with the results from the study of the correlations, we are justified to estimate that at  
$\beta_R=0.155(2)$ a second order phase transition point should be expected.

\section{Conclusions}

We believe that we have serious evidence that the five--dimensional Abelian Higgs model 
with strong coupled interactions along the fifth (transverse) direction  
reveals a four--dimensional dynamics with  broken gauge symmetry. This occurs  via a second order 
phase transition.  
The existence of the layered phase  can be considered as a realisation for the  
localisation of the gauge and scalar fields for models defined in a higher 
dimensional space with the extra dimensions being warped. 
Although the lattice volumes and the computer power available is not  conclusive for the second order 
critical point  (so that the calculation of critical exponents is out of consideration for the 
moment), our results provide an estimate  
for the value of the Higgs self coupling at which  the line of the 
first order transition line  ends in a second order transition point along the four--dimensional 
space.

\section{Acknowledgements}

We thank  A. Kehagias and G. Koutsoumbas for reading and comments on the manuscript.
We  also wish to thank K. Bachas, M. Giovannini for useful discussions. 
Thanks are also due to the Computer Center of the NTUA for the computer time allocated.
The work of P.D. is partially supported by "Thales" project of NTUA and the "Pythagoras"
project of the Greek Ministry of Education.

\end{document}

%% file: macros.tex
\newcommand{\ba}{\begin{array}}
\newcommand{\ea}{\end{array}}

\newcommand{\Dslash}{\relax{\kern+.25em / \kern-.70em D}}

\newcommand{\Real}{\relax{\mathsf{\Gamma\kern-.35em R}}}
\newcommand{\Int}{\relax{\mathsf{Z\kern-.40em Z}}}

\newcommand{\gbar}{\kern1pt\overline{\kern-1pt g\kern-0pt}\kern1pt}
\newcommand{\mbar}{\kern2pt\overline{\kern-1pt m\kern-1pt}\kern1pt}
\newcommand{\obar}[1]{\kern3pt\overline{\kern-2pt #1\kern-0pt}\kern1pt}





%% file: Hlay_v3.bbl
\begin{thebibliography}{99}
\bibitem{funiel} Y.K. Fu and H.B. Nielsen, Nucl. Phys. {\bf B
236}, 167 (1984); Nucl. Phys. {\bf B236}, 127 (1985).
\bibitem{rab} D. Berman and E. Rabinovici, Phys. Lett. {\bf B157} 292  (1985).
\bibitem{stam} C.P. Korthals-Altes, S. Nicolis and J. Prades,
Phys. Lett. {\bf B316} 339 (1993) [hep-lat/9306017]; 
A. Hulsebos, C.P. Korthals-Altes and S. Nicolis, Nucl. Phys. {\bf
B450} 437 (1995) [hep-th/9406003]. 
\bibitem{dim1} P. Dimopoulos, K. Farakos, A. Kehagias and G. Koutsoumbas, 
Nucl.Phys.{\bf B617}:237-252,2001  [hep-th/0007079]; 
\bibitem{ejiri} S. Ejiri, J. Kubo and M. Murata, Phys.Rev.{\bf D62}:105025,(2000) [hep-ph/0006217];
S. Ejiri, S. Fujimoto and J. Kubo, Phys.Rev.{\bf D66}:036002,(2002) [hep-lat/0204022].
\bibitem{farakos} K. Farakos, P. de Forcrand, C.P. Korthals Altes, M. Laine and M. Vettorazzo, 
\bibitem{dim2} P. Dimopoulos, K. Farakos, G. Koutsoumbas, C.P.
Korthals-Altes, S. Nicolis,  JHEP {\bf 02}(005)
(2001) [hep-lat/0012028]; 
\bibitem{dim3} P. Dimopoulos, K. Farakos and  S. Nicolis,
Eur.Phys.J.{\bf C24}:287-296,2002,  [hep-lat/0105014]. 
\bibitem{rummu} M.Laine, H.B.Meyer, K.Rummukainen and M.Shaposhnikov, {\it 
Effective Gauge Theories on Domain Walls via Bulk Confinement?} [hep-ph/0404058].
\bibitem{dvali} G.R. Dvali and M.A. Shifman, Phys.Lett.{\bf B396}:64 (1997), 
Erratum-ibid.{\bf B407}:452 (1997) [hep-th/9612128]. 
\bibitem{dim4} P. Dimopoulos, K. Farakos and G. Koutsoumbas, Phys.Rev.{\bf D65}:074505,2002 [hep-lat/0111047].
\bibitem{dienes} K.R. Dienes, E. Dudas and T. Gharghetta, Nucl. Phys. {\bf B537} (1999) 47, [hep-ph/9806292].
\bibitem{alonso} J.L.Alonso et. al.  Nucl.Phys. {\bf B405}  (1993) 574, [hep-lat/9210014]
\bibitem{rummu_2} K.Kajantie, M.Laine, K.Rummukainen and M.Shaposhnikov,
Phys.Rev.Lett. {\bf 77}  2887 (1996)  [hep-ph/9605288]; 
K.Rummukainen, M.Tsypin, K.Kajantie, M.Laine and M.Shaposhnikov, Nucl. Phys. {\bf B532} 283 (1998)
[hep-lat/9805013] 
\bibitem{tepper} A. Hart, O. Philipsen, J.D. Stack and M. Teper, 
Phys.Lett. {\bf B396}  217 (1997)[hep-lat/9612021]. 
\bibitem{randall} L. Randall and R. Sundrum, Phys. Rev. Lett. {\bf 83} (1999) 4690, [hep-th/9906064];
L. Randall and R. Sundrum, Phys. Rev. Lett. {\bf 83} (1999) 3370, [hep-th/9905221]. 
\bibitem{kehagias} A. Kehagias, Phys. Lett. {\bf B469} (1999) 123, [hep-th/9906204];
A. Kehagias and K. Tamvakis, Phys. Lett. {\bf B504} (2001) 38, [hep-th/0010112].
\bibitem{giovannini} M. Giovannini, Phys. Rev. {\bf D64} (2001) 124004, [hep-th/0107233];
Phys.Rev. {\bf D65} (2002) 124019, [hep-th/0204235].
\bibitem{Shaposh} M.Shaposhnikov and P.Tinyakov, Phys.Lett. {\bf B515} (2001) 442-446, [hep-th/0102161];
M.Laine, H.B.Meyer, K.Rummukainen and M.Shaposhnikov, JHEP {\bf 0301} (2003) 068 [hep-ph/0211149].
\bibitem{Rubakov}V.A. Rubakov, Usp.Fiz.Nauk {\bf171}  913-938 (2001),   [hep-ph/0104152]
\end{thebibliography}
